\documentclass[12pt]{article}

\usepackage[totalheight = 23cm, totalwidth = 17cm]{geometry}
\usepackage{amssymb,amsmath,amsfonts,amsbsy,graphicx}
\usepackage{epsf}
\usepackage{epstopdf}

\def\mathbi#1{\textbf{\em #1}}

\newcommand{\mpl}{m_{\rm Pl}}
\newcommand{\fnl}{f_{\rm NL}}
\newcommand{\calB}{{\cal B}}
\newcommand{\calC}{{\cal C}}
\newcommand{\calM}{{\cal M}}
\newcommand{\calP}{{\cal P}}
\newcommand{\calR}{{\cal R}}
\newcommand{\calS}{{\cal S}}
\usepackage{color}

\begin{document}

\begin{titlepage}

\rightline{\footnotesize{APCTP-Pre2014-008}} \vspace{-0.2cm}

\begin{center}

\vskip 1.0cm

\Large{\bf Scale-dependent hemispherical asymmetry \\ from general initial state during inflation}

\vskip 1.0cm

\large{
Hassan Firouzjahi$^a$,
\hspace{0.1cm}
Jinn-Ouk Gong$^{b,c}$
\hspace{0.1cm} and \hspace{0.1cm}
Mohammad Hossein Namjoo$^{a}$
}

\vskip 0.5cm

\small{\it 
$^{a}$School of Astronomy, Institute for Research in Fundamental Sciences (IPM) \\
P.~O.~Box 19395-5531, Tehran, Iran
\\
$^{b}$Asia Pacific Center for Theoretical Physics, Pohang 790-784, Korea 
\\
$^{c}$Department of Physics, Postech, Pohang 790-784, Korea
}

\vskip 1.2cm

\end{center}

\begin{abstract}

We consider a general initial state for inflation as the mechanism for generating
scale-dependent hemispherical asymmetry. An observable scale-dependent non-Gaussianity is generated  that leads to  observable hemispherical asymmetry from the super-horizon long mode modulation.  We show that the amplitude of dipole asymmetry falls off exponentially on small
angular scales which can address the absence of dipole asymmetry at these scales.
 In addition, depending on the nature of non-vaccum initial state, the amplitude of the dipole asymmetry has oscillatory features which can be detected in a careful CMB map analysis. Furthermore, we show that the non-vacuum initial state provides a natural mechanism for  enhancing the super horizon long mode  perturbation as required to generate the dipole asymmetry.

\end{abstract}

\end{titlepage}

\setcounter{page}{0}
\newpage
\setcounter{page}{1}

\section{Introduction}

There are indications of dipole asymmetry in the cosmic microwave background (CMB) map as reported by PLANCK~\cite{Ade:2013nlj,Ade:2013zuv}, see also~\cite{other-aniso}. 
There are debates on how significant these asymmetries are as were critically reviewed in WMAP reports~\cite{Bennett:2010jb}. For example, 
there were earlier suggestions of the detection of quadrupole asymmetry in WMAP data but later analysis demonstrated that the systematic errors can address the apparent source of quadrupole asymmetry~\cite{wmapaniso-crit}. Therefore, while looking for theoretical explanations of the source of dipole asymmetry, one has to allow for the possibility that there may exist systematic errors which can cause the dipole asymmetry on the CMB map. It is expected that the second year PLANCK data with new results on polarizations and more data analysis can provide further clues about the reality of the current observed dipole asymmetry.

Having the above discussion in mind, we consider the dipole asymmetry as an observed and real phenomenon. A phenomenologically useful way to parameterize the observed hemispherical asymmetry in the 
comoving curvature perturbation power spectrum $\calP_\calR(k)$ on the CMB is 
\begin{equation}
\label{P-asym}
\calP^{1/2}_\calR(k,\mathbi{x}) = \left[ 1+ A(k) \frac{\hat{\mathbi{p}}\cdot\mathbi{x}_\text{cmb}}{x_\text{cmb}} \right] \calP^{1/2}_\text{iso}(k) \, ,
\end{equation}
where $\calP_\text{iso}$ is the isotropic power spectrum and $A(k)$ denotes the amplitude of the dipole asymmetry. In addition, the direction of anisotropy is shown by $\hat{\mathbi{p}}$ and $x_\text{cmb}$ represents the comoving distance to the surface of last scattering. In  the simplest with a scale free asymmetry, i.e. $A(k)=A$, the above asymmetric power spectrum translates to the following CMB temperature anisotropy
\begin{equation} 
\Delta T(\hat{\mathbi{n}}) = \Delta T_\text{iso} \left( 1+A \hat{\mathbi{p}}\cdot\hat{\mathbi{n}} \right) \, ,
\end{equation}
in which $\hat{\mathbi{n}}$ is the direction of observation.
The recent data from PLANCK observations indicates a detection of dipole asymmetry with the amplitude  $A =0.072 \pm 0.022$  for $\ell < 64$.  In addition,  the best fit for the anisotropy direction is $(l,b)=(227,-27)$~\cite{Ade:2013nlj}. Note that this hemispherical asymmetry does not survive at smaller scales and it practically vanishes at $\ell > 600$. There are also tight constraints on such asymmetry at sub-CMB scales from e.g. quasar observations~\cite{Hirata:2009ar}, etc. This suggests a scale dependent asymmetry $A(k)$. Our main goal in this article is to propose a mechanism fo such scale-dependent  asymmetry.

One particular theoretical suggestion for the source of dipole asymmetry is the idea of super-horizon long mode modulation~\cite{aniso-longmode,Erickcek:2008jp,Erickcek:2009at}. Historically, the effects of super-horizon modes on cosmological parameters were also investigated by Grishchuk and  Zel'dovich~\cite{Grishchuk1978}. In this view, our observed universe may be part of a very large ``super universe''. For some unknown reasons, it is possible that one long mode is excited in this super universe with an amplitude much larger than the observed COBE normalization $\calR \sim 10^{-5}$ on the CMB scales. An observer in our observable patch does not have access to the wave-like nature of this perturbation. Instead, for an observer localized in our Hubble patch this very long mode causes small changes in the background quantities such as the value of the background inflaton field or the total number of inflationary $e$-folds. This in turn yields a dipole asymmetry which may be viewed as the source of observed dipole asymmetry. This idea has attracted considerable interests~\cite{Lyth:2013vha,Namjoo:2013fka,Abolhasani:2013vaa,many}, see also \cite{Aslanyan:2013jwa}

In order for the super-horizon long mode modulation to generate observable dipole asymmetry one requires a large coupling between the large mode, with wavenumber denoted by $k_L$, and the smaller 
CMB scale modes: non-Gaussianity in squeezed limit, i.e. in the limit in which one mode has a wavelength much larger than the other two. This rings the bells in many ways. First, from the recent PLANCK observations there is no detection of the primordial non-Gaussianity~\cite{Ade:2013ydc}, so a priori it is non-trivial how one may get large dipole amplitude. Secondly, for single field models of inflation, the amplitude of non-Gaussianity in the squeezed limit, here denoted by $\fnl$, is subject to the non-Gaussianity consistency condition~\cite{Maldacena:2002vr,Creminelli:2004yq}  $\fnl \sim n_\calR -1$ in which $n_\calR$ is  the tilt of $\calP_\calR$. With $n_\calR \simeq 0.96$ from the CMB observations,  $\fnl$ in standard single field inflation is too small to generate the observed dipole asymmetry. This have been demonstrated in a model independent way in~\cite{Namjoo:2013fka}.

To bypass this difficulty, one has to break the non-Gaussianity 
consistency relation. The simplest approach is to employ multiple field models of inflation where non-negligible $\fnl$ can be generated~\cite{Erickcek:2009at,Lyth:2013vha}. Another approach is to consider non-attractor models, where the curvature perturbation is not frozen on large scales so that $\fnl$ becomes large~\cite{nonattractor}. Once the system reaches the attractor limit $\fnl$ becomes negligible. This built-in scale-dependence of $\fnl$ helps to quickly diminish the dipole asymmetry on sub-CMB scales such as on quasar scales~\cite{Hirata:2009ar,Flender:2013jja}, see also \cite{Notari:2013iva}. 
Alternatively, if one starts with the non-vacuum initial states during inflation, large $\fnl$ can be generated in the squeezed limit~\cite{ Chen:2006nt,Holman:2007na,nBD-fNL, Agullo:2010ws, Ganc:2011dy}. Usually it is assumed that inflation starts with a Minkowski or Bunch-Davies (BD) initial condition. This is the case if inflation lasts for a long period in the past. However, it is possible that universe may experience a transient period before inflation starts, or there were phase transitions or particle creations at the start of inflation. Then, it is natural to consider a generalized, non-BD initial condition. In this article we consider how starting with a non-BD initial condition can help to produce large scale-dependent dipole asymmetry. In addition, we will show that non-BD initial condition can induce an enhancement on the amplitude of super-horizon long mode which is responsible for the asymmetry. This helps to obtain observable asymmetry at the CMB scales. Note, however, that arbitrary non-BD states may not be compatible with the current observations and, even if so, may not lead to observational non-Gaussian signatures~\cite{no-nBD}. We do not aim at constructing explicit non-BD states that survive observational and theoretical constraints, but present generic results that can be applied to viable non-BD cases.

The rest of this article is organized as follows. In Section~\ref{sec:dipole-fNL} we review the results obtained in~\cite{Namjoo:2013fka,Abolhasani:2013vaa} relating $A$ to $\fnl$ for a generic single field model. In Section~\ref{sec:ng} we study a generic single field model of inflation  with a non-BD initial condition which can yield large $\fnl$ with non-trivial shape. In Section~\ref{CMB-constraints} we present the CMB constraints and predictions for the amplitude 
and the shape of dipole asymmetry with a non-BD initial condition. 
In Section~\ref{sec:conc} we conclude.

\section{Hemispherical asymmetry and  non-Gaussianity}
\label{sec:dipole-fNL}

In this section, we briefly review the relation between the amplitude of the dipole asymmetry and non-linear parameter $\fnl$. We mostly follow~\cite{Namjoo:2013fka,Abolhasani:2013vaa}.

Our aim is to parametrize the effects of the long mode $k_L$ on the CMB modes $k$ as given in (\ref{P-asym}). The fractional change due to the long wavelength modulation on the power spectrum is then
\begin{equation}
\label{grad}
\frac{\nabla P_\calR}{P_\calR} = \frac{2A\hat{\mathbi{p}}}{x_\mathrm{cmb}} \, ,
\end{equation}
where $P_\calR(k) \equiv 2\pi^2\calP_\calR(k)/k^3$. Meanwhile, considering the three-point correlation function in the squeezed limit, with $\mathbi{k}_1$ and $\mathbi{k}_2$ denoting CMB scale modes, gives
\begin{align}
\label{rescale}
\Big \langle \calR(\mathbi{k}_L) \calR(\mathbi{k}_1) \calR(\mathbi{k}_2) \Big \rangle
& \approx
\Big{\langle} \calR(\mathbi{k}_L) \Big{\langle} \calR(\mathbi{k}_1)\calR(\mathbi{k}_2) \Big\rangle_{\calR(\mathbi{k}_L)}  \Big\rangle 
\nonumber\\
& \approx
\Big \langle \calR(\mathbi{k}_L) \left[ \calR(\mathbi{k}_L) \dfrac{\partial}{\partial \calR(\mathbi{k}_L)} \Big \langle \calR(\mathbi{k}_1)\calR(\mathbi{k}_2) \Big \rangle \large \Big\vert_{\calR(\mathbi{k}_L)=0}  \right]
\Big\rangle \, .
\end{align}
This is the key equation for our analysis below. Intuitively this equation means  that, as long as the small scales perturbations are concerned, the effect of a large scale perturbation is just a rescaling of the background  scale factor. This is motivated from the fact that  the  small scale perturbations can not probe the spatial variations associated with the long wavelength mode. 
This is the case if the long wavelength mode is larger  than the Hubble patch of the CMB scale modes. One important assumption for the validity of the above relation is that the whole three-point correlation is generated when the long mode has left the horizon, i.e. after it becomes classical.  That is, there is no correlation between modes when the long mode and hence the small modes are all deep inside the horizon. This latter condition may be violated in models of non-BD initial conditions in which there is a finite ultraviolet (UV) initial time cutoff in the model. As a result, in order for the above relation to be valid, we consider non-BD models in which the initial time cutoff is set to infinity, corresponding to the case in which all modes oscillate rapidly deep inside the horizon, canceling all sub-horizon correlations.
Another key assumption for the validity of (\ref{rescale}) is that there is only one degree of freedom encoded in $\calR$, so there is no other source of perturbation.  This line of thought was used in \cite{Creminelli:2004yq,squeezedbi} to calculate the amplitude of the non-Gaussianity in squeezed limit and to check the non-Gaussianity consistency condition for single field inflation models.

Equipped with (\ref{rescale}) we can relate the amplitude of dipole asymmetry $A$ to the non-linear parameter $\fnl$ as follows. Let us define the bispectrum $B_\calR(k_1,k_2,k_3)$ as
\begin{equation}
\left\langle \calR(\mathbi{k}_1)\calR(\mathbi{k}_2)\calR(\mathbi{k}_3) \right\rangle \equiv (2\pi)^3 \delta^{(3)}(\mathbi{k}_1+\mathbi{k}_2+\mathbi{k}_3) B_\calR(\mathbi{k}_1,\mathbi{k}_2,\mathbi{k}_3) \, .
\end{equation}
Comparing with(\ref{rescale}), we obtain
\begin{equation}
\label{B-eq}
B_\calR(\mathbi{k}_1,\mathbi{k}_2,\mathbi{k}_L)  = P_\calR(k_L) \frac{\partial P_\calR(k_1)}{\partial \calR_L} \, ,
\end{equation}
where $\calR_L \equiv \calR(\mathbi{k}_L)$.
The non-linear parameter $\fnl$ in the squeezed limit $k_1 \approx k_2 \gg k_L$ is defined via
\begin{equation}
\label{fNL-def}
\fnl = \lim_{k_L \ll k_1,k_2} \frac{5}{12} \frac{B_\calR(\mathbi{k}_1,\mathbi{k}_2,\mathbi{k}_L)}{P_\calR(k_1) P_\calR(k_L)} \, .
\end{equation}
Combining (\ref{B-eq}) and (\ref{fNL-def}) we obtain 
\begin{equation}
\frac{1}{P_\calR(k)} \frac{\partial P_\calR(k)}{\partial \calR_L} = \frac{12}{5} \fnl \, ,
\end{equation}
in which it is understood that $k=k_1=k_2$ represents  to the (small) CMB-scale perturbations. Since the effects of the long mode modulation is considered as a directional modification of the power spectrum, it is convenient to write the above equation as
\begin{equation}
\frac{\nabla P_\calR}{P_\calR} = \frac{12}{5} \fnl \nabla \calR_L \, .
\end{equation}
Comparing this equation with (\ref{grad}), we finally obtain the  following formula for $A$:
\begin{equation}
\label{A-eq}
A = \frac{6}{5} \fnl x_\mathrm{cmb}  \left| \nabla \calR_L \right| \, .
\end{equation}

We can proceed further under the assumption that there exists a {\em single} large super-horizon model $\calR_L$ with the amplitude $\calP_L$ and the comoving wavenumber $k_L$,
\begin{equation}
\label{RL}
\calR_L = \calP_L^{1/2}\sin (\mathbi{k}_L\cdot\mathbi{x}) \, .
\end{equation}
Then, we have
\begin{equation}
\label{gradR}
\left| \nabla \calR_L \right| =  k_L \left| \calR_L \right| \approx k_L \calP_L^{1/2} \, .
\end{equation}
Now using this into (\ref{A-eq}), we obtain 
\begin{equation}
\label{AfNL}
A(k) = \frac{6}{5} \fnl k_L x_\mathrm{cmb}  \, \calP_L^{1/2} \, .
\end{equation}
This is the consistency condition obtained in~\cite{Namjoo:2013fka,Abolhasani:2013vaa}.
Note that this formula applies for all models of inflation in which the curvature perturbation has a single source such as in single field models. This relation was first obtained in~\cite{Lyth:2013vha} for the special case of curvaton model.

One comment is in order before we discuss the consistency (\ref{AfNL}).
In writing (\ref{RL}), we assume that our observed universe is part of a much larger universe, i.e. a large box universe 
in the view of~\cite{Lyth:2007jh} with the large super-horizon mode $\calR_L$ being superimposed in our observable universe.
In this view, the quantum fluctuations of $\calR$ associated with the small-scale perturbations inside this large box, $k \gg k_L$, are treated as random statistical variables. In this picture the size of our observed Universe is given by $H_0^{-1}$ in which $H_0$ is the current Hubble constant but the long mode which causes the modulation has the wavelength $\lambda_L \gg H_0^{-1}$. Suppose, for our small CMB-scale modes $k$,  we work in the Fourier space with the volume $V$.  For this picture to be consistent, the volume of the Fourier space  should be bigger than $H_0^{-1}$ but smaller than $\lambda_L$ so the following hierarchy is at work:
\begin{equation}
H_0^{-3} < V \ll k_L^{-3} \, .
\end{equation}

In order to use (\ref{AfNL}) for the practical purpose, one has to eliminate the combination 
$k_Lx_{\mathrm{cmb}}  \, \calP_L^{1/2}$ using observational constraints. The simplest 
constraint is  $ \calR_L^2  \simeq   \calP_L \lesssim 1$. This is necessary in order  for perturbations to be under control. Secondly, one has to check the constraints from the quadrupole $Q_2$ and octupole $Q_3$~\cite{Erickcek:2008jp}. 
It turns out that the quadrupole imposes a tighter constraint~\cite{Erickcek:2008jp,Lyth:2013vha} 
and after eliminating the combination $k_L x_\mathrm{cmb} \, \calP_L^{1/2}$ using these constraints one obtain~\cite{Lyth:2013vha}
\begin{equation}
\label{A-upper-lyth}
 | A |  \lesssim 0.02 \, \left| \fnl \right|^{1/2}  \, .
\end{equation}
In order to obtain  an asymmetry consistent with the PLANCK observation we need $\vert A \vert =0.07 \pm 0.02$. For single field slow roll models of inflation 
we have $\fnl \sim n_\calR-1 \sim -0.04$ and in turn $A \sim 10^{-3}$ from (\ref{A-upper-lyth}), which is too small to explain the observed hemispherical asymmetry. 
As a result, (\ref{A-upper-lyth}) indicates that
in order to have large observable hemispherical asymmetry one has to violate the non-Gaussianity consistency condition. There are two known mechanism for this. One 
is to consider non-attractor models in which the curvature perturbation is evolving on super-horizon scale. This results in large $\fnl$ which in turn yields observable value for $A$ as  studied 
in \cite{Namjoo:2013fka}. The other
is to consider a non-vacuum initial state, i.e. a non-BD initial condition. It is known that for non-BD initial condition one can obtain large $\fnl$. In addition, the non-Gaussianity has a non-trivial shape. 
In the following Sections we study the effects of non-BD initial condition in generating hemispherical asymmetry.

\section{Non-Gaussianity from general initial state}
\label{sec:ng}

In this Section we study single field models of inflation with a non-BD initial condition which
can violate the non-Gaussianity consistency condition and generate large dipole asymmetry.

We consider a broad class of models of inflation with a non-standard kinetic energy.  
When building an inflation model, usually the kinetic term is canonically normalized and the potential is set free. This is however not necessarily the only possibility. In many models originated from e.g. string theory, we expect corrections to the canonical kinetic term. These corrections include not only first derivatives $\partial_\mu\phi$ but also higher order ones, such as $\Box\phi$ which should however be suppressed by powers of the UV cutoff scale like the Planck mass $\mpl$. Thus, we consider the matter Lagrangian of the general form
\begin{equation}
S = \int d^4x \sqrt{-g} P(X,\phi) \, ,
\end{equation}
with $X \equiv -g^{\mu\nu}\partial_\mu\phi\partial_\nu\phi/2$. For canonical case, $P=X-V$.  This action includes a large class of models and situations, such as $k$-inflation \cite{k-inf}, 
Dirac-Born-Infeld inflation \cite{Chen:2006nt,dbi}, effective single field model with heavy fields being integrated out~\cite{int-out} and so on.

The quadratic action of the curvature perturbation $\calR$ is then specified by defining the ``speed of sound'' $c_s$ additionally,
\begin{equation}
c_s^2 \equiv \frac{P_X}{P_X+2XP_{XX}} \, ,
\end{equation}
where $P_X \equiv dP/dX$, and is written as
\begin{equation}\label{S2}
S = \int d^4x a^3\mpl^2\epsilon \left[ \frac{\dot\calR^2}{c_s^2} - \frac{(\nabla\calR)^2}{a^2} \right] \, ,
\end{equation}
in which $a$ is the background scale factor and $\epsilon \equiv - \dot H/H^2$ is the slow-roll parameter.

\subsection{General initial state}

Let us look at the curvature perturbation $\calR$ in Fourier space as a quantum operator,
\begin{equation}
\calR(\eta,\mathbi{k}) = a_\mathbi{k}\widehat\calR_k(\eta) + a_{-\mathbi{k}}^\dag \widehat\calR_k^*(\eta) \, ,
\end{equation}
in which $a_\mathbi{k}$ and $a_\mathbi{k}^\dag$ are the usual annihilation and the creation operators, $\widehat\calR_k(\eta)$ is the mode function of the curvature perturbation and $d\eta = dt/a$ is the conformal time. For the BD initial condition, the mode function solution at leading order in the slow-roll approximation is given by
\begin{equation}
\widehat\calR_{\mathrm{BD}}(k, \eta) = \frac{iH}{\sqrt{4\epsilon c_sk^3}\mpl} (1+ic_sk\eta)e^{-ic_sk\eta} \, .
\end{equation}
A good method to obtain a generalized non-vacuum initial state is to perform a Bogoliubov transformation on the BD vacuum so the mode function is given by
\begin{align}
\label{u-nBD}
\widehat\calR_k(\eta) & = C_k \widehat\calR_{\mathrm{BD}}(k, \eta)  + D_k \widehat\calR_{\mathrm{BD}}(k, \eta)^* 
\nonumber\\
& = \frac{iH}{\sqrt{4\epsilon c_sk^3}\mpl} \left[ C_k(1+ic_sk\eta)e^{-ic_sk\eta} + D_k(1-ic_sk\eta)e^{ic_sk\eta} \right] \, ,
\end{align}
where the coefficients $C_k$ and $D_k$ are subject to the normalization
\begin{equation}
\label{normal}
|C_k|^2 - |D_k|^2 =1  \, .
\end{equation}
In this view, the Minkowski or BD initial condition is given by $C_k=1$ and $D_k=0$.
Furthermore, a non-zero $D_k$ can be viewed as the presence of additional particle state with the number density $N_k$ per momentum interval given by
\begin{equation}
\label{Nk}
N_k = | D_k|^2 \, .
\end{equation}

The important question is what are the physical constraints on the Bogoliubov coefficients $C_k$ and $D_k$. There are some  simple constraints which should be implemented when considering non-BD initial condition \cite{Holman:2007na}. One important requirement is that the total energy density associate with the non-BD fluctuations to be finite. To see this, suppose we interpret the non-BD initial state as the state in which there are particle excitations with the number density $N_k$ given by (\ref{Nk}).  As a result 
the number density of the quanta in the proper unit volume is $ | D_k|^2 d^3 k/( 2 \pi a)^3 $. 
Adding up the energy associated with these modes, their contribution in energy density should remain finite so $\int d^3k \, k{N}_k$ converges. This can be satisfied if 
${ N}_k = {\cal O}(1/k^{4+\delta})$ with $\delta >0$ in the UV region. One can interpret this  as the renormalizability condition. The second, stronger requirement is that the back-reaction from the non-BD excited states do not stop inflation, which implies that 
\begin{equation}
\int d^3k \, k{N}_k  \lesssim \mpl^2 H^2 \, .
\end{equation}
Finally, one should make sure that the non-BD fluctuations do not change the near scale-invariant shape of $\calP_\calR$. 
The change in the tilt of power spectrum, $\delta n_\calR$,  induced from the non-BD fluctuations is \cite{Ganc:2011dy}
\begin{equation}
\delta n_\calR = \frac{d \log(1 + 2  N_k)}{d \log k} \, .
\end{equation}
From the PLANCK data we have $n_\calR \simeq 0.96$ so the change in 
$\delta n_\calR$ can be at most at the order of few percent.

\subsection{Non-linear parameter}

The leading order cubic action for $\calR$ is given by~\cite{Chen:2006nt}
\begin{align}
S_3 = & \int d^4x \left\{ -a^3\left[ \Sigma\left( 1-\frac{1}{c_s^2} \right) +2\lambda \right] \frac{\dot\calR^3}{H^3} + \frac{a^3\epsilon\mpl^2}{c_s^4} \left( \epsilon-3+3c_s^2 \right) \dot\calR^2\calR \right.
\nonumber\\
& \left. \hspace{1.5cm} + \frac{a^3\epsilon\mpl^2}{c_s^2} \left( \epsilon-2s+1-c_s^2 \right) \calR(\nabla\calR)^2 - 2\frac{a\epsilon\mpl^2}{c_s^2}\dot\calR\calR^{,i}\chi_{,i} \right\} \, ,
\end{align}
Here, $\Sigma \equiv \epsilon\mpl^2H^2/c_s^2$, $\lambda \equiv X^2P_{XX}+2X^3P_{XXX}/3$, $s\equiv\dot{c}_s/(Hc_s)$, and $\Delta\chi \equiv a^2\epsilon\dot\calR/c_s^2$. Compared to the BD case \cite{Maldacena:2002vr} the bispectrum is quite complicated, as we have now many combinations of $D_k$'s. However we can obtain more simplified expression if we take the squeezed limit, $k_1 \approx k_2$ and $k_3 \ll k_1,k_2 $ which is actually what we need as demonstrated in (\ref{rescale}). 
After some calculations, using the leading order power spectrum
\begin{equation}
\label{power-nbd}
\calP_\calR = \frac{H^2}{8\pi^2\mpl^2\epsilon c_s}|C_k+D_k|^2 \, ,
\end{equation}
we find
\begin{align}
B_\calR(k_1,k_2,k_3) \underset{k_3\ll k_1,k_2}{\longrightarrow} & 4\pi^4\calP_\calR(k_1)\calP_\calR(k_3) \left[ -3 \left( c_s^2-1+\frac{2c_s^2\lambda}{\Sigma} \right) + \frac{2\epsilon-2s}{c_s^2} + 2\left( 1-\frac{1}{c_s^2} \right) \right] 
\nonumber\\
& \times \frac{1}{k_1^2k_3^4} \frac{\prod_{i=1}^3 (C_i+D_i)}{|C_1+D_1|^2|C_3+D_3|^2} \left\{ \left[ C_1^*D_2^*C_3^* + D_1^*C_2^*C_3^* - (C \leftrightarrow D) \right] + c.c.  \right\} \, ,
\end{align}
where $C_i = C_{k_i}$ and $D_i = D_{k_i}$, and we have neglected sub-leading terms not boosted by $1/k_3$. If we consider the canonical single field case where $c_s=1$ and $\lambda=s=u=0$, we can recover the result with the BD initial condition \cite{Ganc:2011dy}.

From the normalization (\ref{normal}), we can generally parametrize $C_k$ and $D_k$ as
\begin{align}
C_k & = e^{i\alpha_k}\cosh\chi_k \, ,
\\
D_k & = e^{i\beta_k}\sinh\chi_k \, ,
\end{align}
with the phase difference defined by $\theta_k \equiv \alpha_k-\beta_k$.
Setting $C_1=C_2$ and $D_1=D_2$ in the squeezed limit $k_1 \approx k_2$, i.e. $\chi_1 \approx \chi_2$ and $\theta_1 \approx \theta_2$, the shape function of the bispectrum is 
\begin{align}
& \prod_{i=1}^3 (C_i+D_i) \left[ C_1^*D_2^*C_3^* + D_1^*C_2^*C_3^* - (C \leftrightarrow D) \right] + c.c.
\nonumber\\
& = 4\cosh\chi_1\sinh\chi_2 \left[ \left( 2\cosh^2\chi_1-1 \right)\cos\theta_1 + 2\cosh\chi_1\sinh\chi_1 + 2\cosh\chi_3\sinh\chi_3\sin\theta_2\sin\theta_3 \right] \, .
\end{align}
In addition, from (\ref{Nk}), we can write $\sinh\chi_k = \sqrt{N_k}$ and $\cosh\chi_k = \sqrt{N_k+1}$. Now using (\ref{fNL-def}) we obtain our final result for $\fnl$ with the non-BD initial condition as
\begin{align}
\label{k-dep_fNL}
\fnl = & \frac{5}{12} \left[ -3 \left( c_s^2-1+\frac{2c_s^2\lambda}{\Sigma} \right) + \frac{2\epsilon-2s}{c_s^2} + 2\left( 1-\frac{1}{c_s^2} \right) \right] 
\nonumber\\
& \times \frac{k_1}{k_3} \frac{4\sqrt{N_1(N_1+1)} \left[ (2N_1+1)\cos\theta_1 + 2\sqrt{N_1(N_1+1)} + 2\sqrt{N_3(N_3+1)}\sin\theta_1\sin\theta_3 \right]}{\left[ 1+2N_1+2\sqrt{N_1(N_1+1)}\cos\theta_1 \right]\left[ 1+2N_3+2\sqrt{N_3(N_3+1)}\cos\theta_3 \right]} \, .
\end{align}
We can see that even if the slow-roll parameters are nearly constant, still non-trivial scale-dependence of $\fnl$ can follow from the second line of the above equation.

Note that the above results in the squeezed limit have been obtained when we set $k_3 \ll k_1, k_2$ but kept $k_3$ to be large enough to satisfy the limit $-k_3 \eta_0 \gg 1$. This is required for the consistency of our bispectrum analysis. In the {\it exact} squeezed limit in which $k_3 \to 0$ arbitrarily, the above results do not hold and $\fnl$ in this limit reduces to the one predicted by the consistency relation, as indicated in~\cite{Agullo:2010ws, Ganc:2011dy}.

The shape of $\fnl$ presented in (\ref{k-dep_fNL}) is too complicated to be studied in general situations. For our analysis of the CMB constraints in next section we consider some simple and physically well-motivated parameterization of the mode number $N_k$ and the phase $\theta_k$. 
One convenient modeling is \cite{Holman:2007na}
\begin{equation}
\label{Nk-model}
N_k = N_{0} \exp \left( -\frac{k^2}{k_c^2} \right) \, ,
\end{equation}
in which $N_0$ is a number and $k_c$ may be related to the UV cutoff of the theory $M$ at the initial time of inflation $\eta_0\equiv -1/k_0$ via  $k_c = a(\eta_0) M$. 
As for the phase $\theta_k = k \eta_0$, there are two approaches. One option is to take $\eta_0$ to be fixed as a preferred initial time deep in the UV region. This means that $\theta_k$ linearly depends on $k$. The second is to keep $\theta_k$ fixed for all modes. We study the predictions of both options in the next section.

In order for  the back-reaction from the exited non-BD states does not destroy the slow-roll inflation we need~\cite{Holman:2007na} $\sqrt{N_0} \leq \sqrt{\epsilon}\mpl H/M^2$. 
For the effective field theory description of inflation to be valid one requires that  $H < M$. In addition, the cutoff $M$ is natural to be much smaller than $\mpl$. Therefore, one can easily obtain ${N_0} >1$ while satisfying all physical constraints from the non-BD initial conditions. As a an example,  suppose $M \sim 10^{-6}\mpl$, $H \sim 10^{-8}\mpl$ and $\epsilon \sim 10^{-2}$, then $N_0 \lesssim 10^6$.

\begin{figure}
\center
\includegraphics[scale=.8]{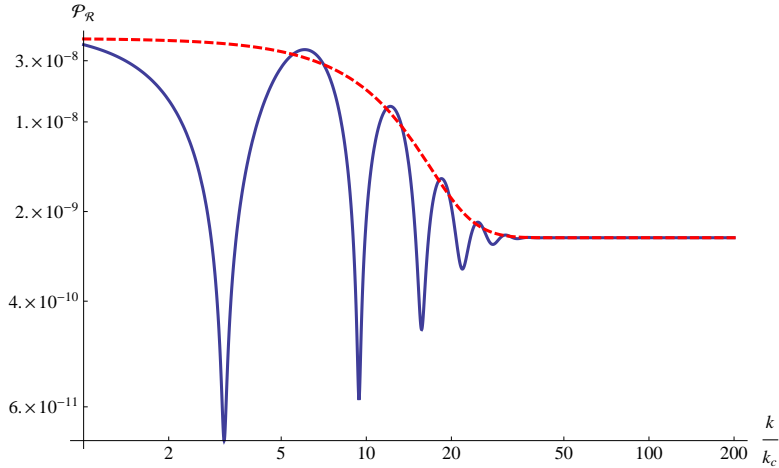}
\caption{The behavior of the power spectrum with non-BD initial condition. For sufficiently small scales we recover the standard scale invariant  power spectrum whereas there is a large enhancement on large super-horizon  scales. The solid line is for $\theta_k =k/k_0$ while the dashed line is for $\theta_k = \pi/4$. We have also set $k_0 = k_c/10 $ and $N_0 =10$.}
\label{power}
\end{figure}

It is worth to mention here an interesting property of the model, i.e. the scale-dependence of the power spectrum as given in (\ref{power-nbd}). 
Note that for $k \gg k_c$, $N_k = \vert D_k \vert^2 \to 0$ resulting in standard power spectrum with BD initial condition. On the other hand, for $k<k_c$ we have a rather large effect from non-BD state enhancing the power spectrum. This interesting observation naturally predicts an enhanced super-horizon long mode, $\calP_L \gg \calP_\calR(k_\text{cmb})$,  which is necessary for generating observable asymmetry~\cite{Lyth:2013vha}. For this aim, one can consider the cutoff scale such that $k_L<k_c  \ll k_\text{cmb}$ so one restores the scale invariance of power spectrum at CMB scales while, at the same time, obtains an enhancement for long mode $\calR_L$. See Figure~\ref{power} for an illustration of this prediction of the model.  Note that the scale-dependence in power spectrum as well as in non-linearity has the same origin, i.e. the specific scale-dependence of $N_k$ in~\eqref{Nk-model}. As we shall see in the next section, the induced scale-dependence in  $\fnl$  from $N_k$ also leads to a scale-dependent hemispherical asymmetry.

\section{CMB analysis and constraints}
\label{CMB-constraints}

In Section \ref{sec:dipole-fNL} we have obtained the relation between $\fnl$ and the amplitude
of dipole asymmetry. Now we consider the CMB constraints on the model parameters. 
Our main goal is to  check 
whether or not the amplitude of asymmetry decays on small scales in real space. As mentioned before, there are strong observational constraints from the sub-CMB scales  on the amplitude of dipole asymmetry.
For example from the absence of dipole asymmetry in quasars number counts \cite{Hirata:2009ar} one concludes that  $ A  \lesssim 10 ^{-2}$ at the scales $k \sim$ Mpc$^{-1}$. In addition, investigating the CMB map on small scales indicates no dipole asymmetry with the upper bound $A \lesssim 10^{-3}$
for $\ell > 600$. Therefore, any theory which can predict dipole asymmetry on $\ell < 64$ as observed from
PLANCK data should also provide a mechanism for the amplitude of dipole asymmetry to decay quickly on smaller angular scales. This calls for a non-trivial scale-dependent $\fnl$.
Our main goal here therefore is to see  whether or not the non-trivial $\fnl$ obtained in~(\ref{k-dep_fNL}) has the right properties to give large dipole amplitude on $\ell < 64$ while decaying rapidly for $\ell > 600$.

It is convenient to separate the $k$-dependence in $\fnl$ from its overall amplitude. Assuming that the slow-roll parameters do not have non-trivial scale dependence, we may define the amplitude $\calM$ and the shape function $\calS$ such that $\fnl \equiv \calM\calS$, with $\calM$ and $\calS$ corresponding to the first and second line of (\ref{k-dep_fNL}), respectively. Then (\ref{A-eq}) can be written as
\begin{equation}
A(k) \equiv \bar{A} \, \calS \, .
\end{equation}
With this convention, in the most conservative case with $\calP_L^{1/2} \simeq 10^{-5}$ and assuming that $k_L$ corresponds to a scale not much larger than the CMB scale, we have $\bar A \sim 10^{-5}/c_s^2$ so one can enhance $A$
by reducing the sound speed. As we shall see below, in order to obtain observable dipole
asymmetry we need $\bar A \sim 10^{-2}-10^{-1}$. This can easily be obtained if we take $c_s^2 \sim 10^{-3}-10^{-2}$
which is consistent with the PLANCK constraint on non-BD initial condition which generate large
non-Gaussianity in folded region as well as in squeezed limit~\cite{Ade:2013ydc}. It is worthwhile to note that we are not forced to satisfy the tight constraints on local non-Gaussianity since our model predicts a quite different shape, though both local and non-BD non-Gaussianities have a peak at squeezed limit.  That is why we do not need a largely enhanced long mode, in contrast to previous models with local non-Gaussianity. However,  as another interesting possibility one can set $c_s$ to a larger value, so that $\fnl$ is reduced, but assume a large amplitude for $\calR_L$. As we have already discussed, this enhancement can easily happen in our model due to the specific properties for non-BD initial conditions.

In order to investigate the effects of the scale-dependence of $\fnl$ in the CMB power spectrum, 
it is useful to consider from (\ref{P-asym}) the extremum power spectrum in the presence of the asymmetry,
\begin{equation}\label{PR-ext}
\calP_\text{ext}^{1/2} = \left[ 1+A(k) \right] \calP_0^{1/2} \, ,
\end{equation}  
which happens when the direction of observation is aligned to $\mathbi{k}_L$. 
From this power spectrum one can compute the observable CMB power spectrum $\calC_\ell$. If the scale-dependence of the asymmetry is appropriate from observational point of view, the deviation of $\calC_\ell$ from the standard $\Lambda$CDM case must decay for sufficiently large $\ell$. In the presence of a hemispherical asymmetry, after computing $\calC_\ell$ from (\ref{PR-ext}), we can decompose the total power spectrum into two pieces
\begin{equation} 
\calC_\ell= \calC_\ell^0 + \Delta \calC_\ell \, ,
\end{equation} 
where $\Delta \calC_\ell$ is the correction to the power spectrum due to the asymmetry and $\calC_\ell^0$ is the $\Lambda$CDM power spectrum. The scale-dependent amplitude of asymmetry on the CMB temperature anisotropy can be parametrized by~\cite{Erickcek:2009at}
\begin{equation} 
K_\ell \equiv \dfrac{\Delta \calC_\ell}{\calC_\ell^0} \, .
\end{equation}
By this definition, we can also define the effective amplitude of asymmetry on scales in the range $2<\ell <\ell_\text{max}=64$,
\begin{equation} 
\calB \equiv \dfrac12 \sum_{\ell =2}^{\ell_\text{max}} \frac{(2 \ell+1)}{(\ell_\text{max}-1) (\ell_\text{max}+1)} K_\ell \, ,
\end{equation} 
where the prefactor comes from the fact that there are $2\ell+1$ independent modes for each $\ell$. 
With this convention, $\calB$ reduces to the scale independent amplitude $A$ if $K_\ell$ is scale independent, consistent with the parameterization of asymmetry in Fourier space
\begin{equation} 
\Delta T({\mathbi{k}}) = \Delta T_0(k) \left( 1 + \calB \frac{\hat{\mathbi{p}}\cdot\mathbi{x}}{x_\text{cmb}} \right) \, . 
\end{equation}

With these discussions, now we present the CMB predictions of the model. We employ the phenomenological parameterization  for $N_k$ as given in (\ref{Nk-model}).  As for the phase $\theta_k$, as we already discussed, we consider both options, i.e. the constant phase and linear scale dependence.

\begin{figure}
\center 
\includegraphics[width=8cm]{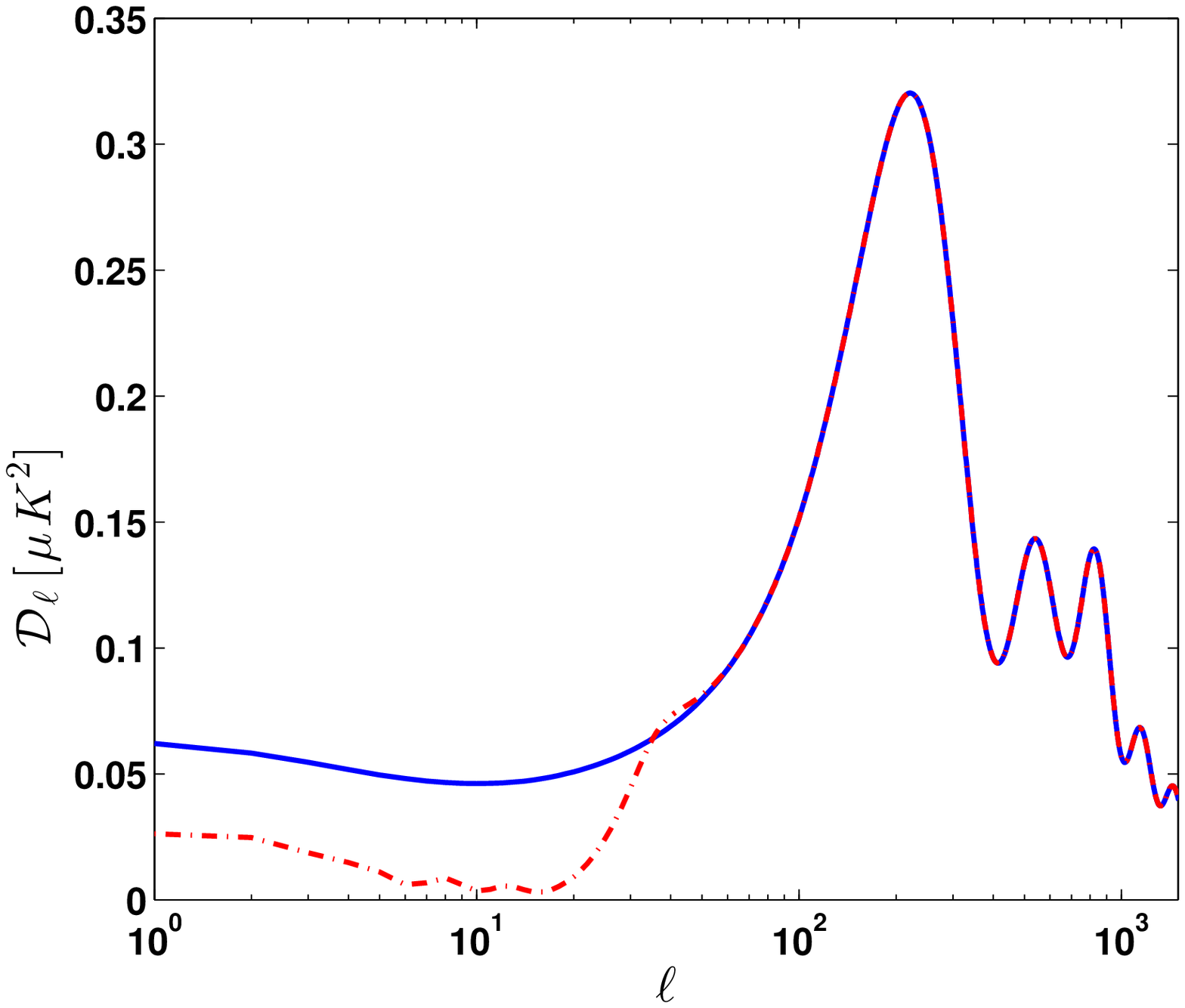}
\includegraphics[width=8cm]{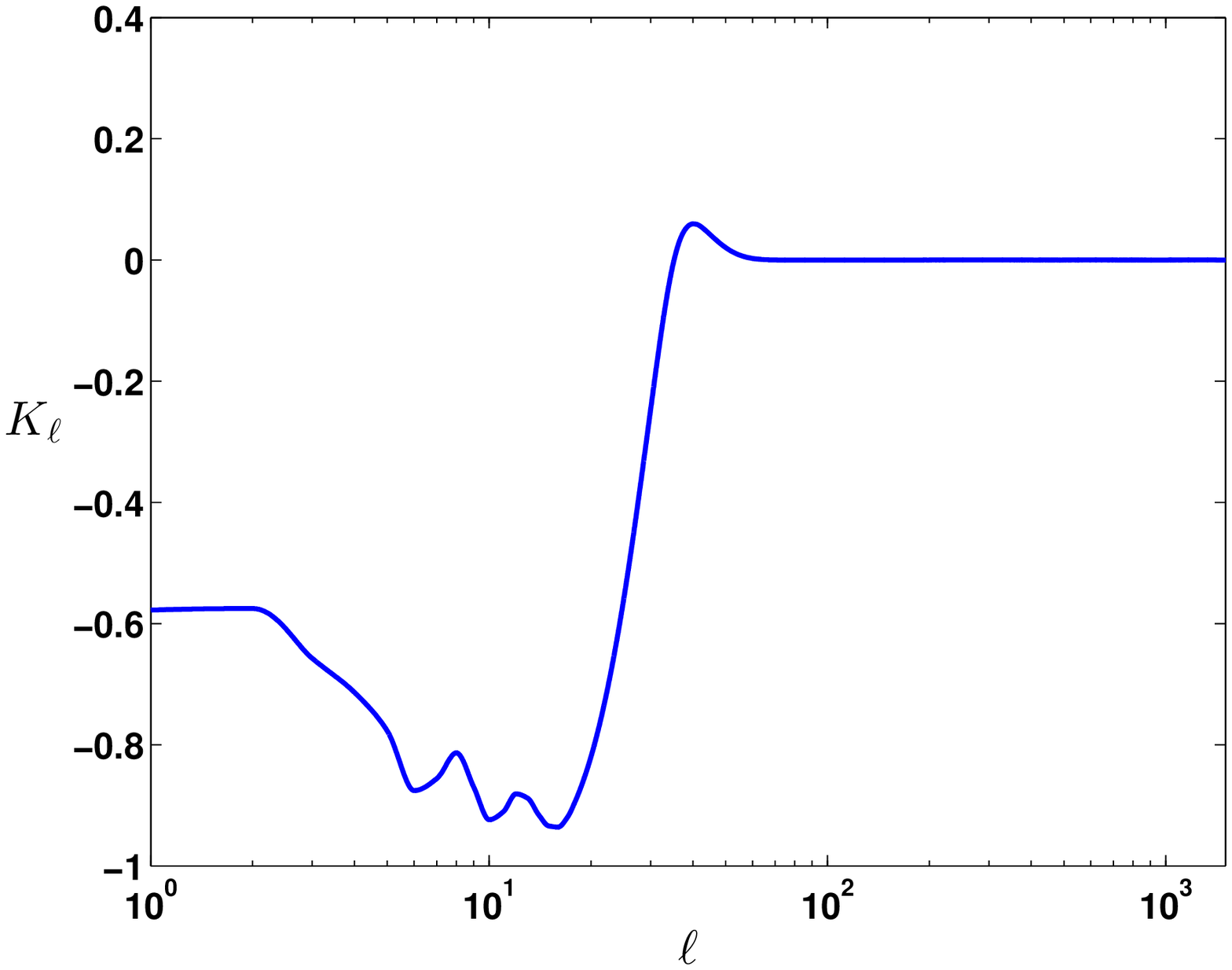}
\caption{Plots for $ \calC_\ell^0 $ (solid curve)  and $ C_\ell + \Delta \calC_\ell $ (dotted-dashed curve).  Here we have  set $N_0=10$, $\bar A=-0.095$ and for the momenta in Mpc$^{-1}$ units: $k_L = 0.00006$, $k_c = 0.0013$ and $k_0= 0.00001$. The effective amplitude of asymmetry is $\calB \simeq 0.0722$ to match the PLANCK observation on large CMB scales. 
As can be seen from the behavior of $K_\ell$ on the right panel, the amplitude of dipole asymmetry rapidly goes to zero for large $\ell$ while it shows oscillatory features for small $\ell$. 
}
\label{case1}
\end{figure}

\begin{figure}
\center 
\includegraphics[width=8cm]{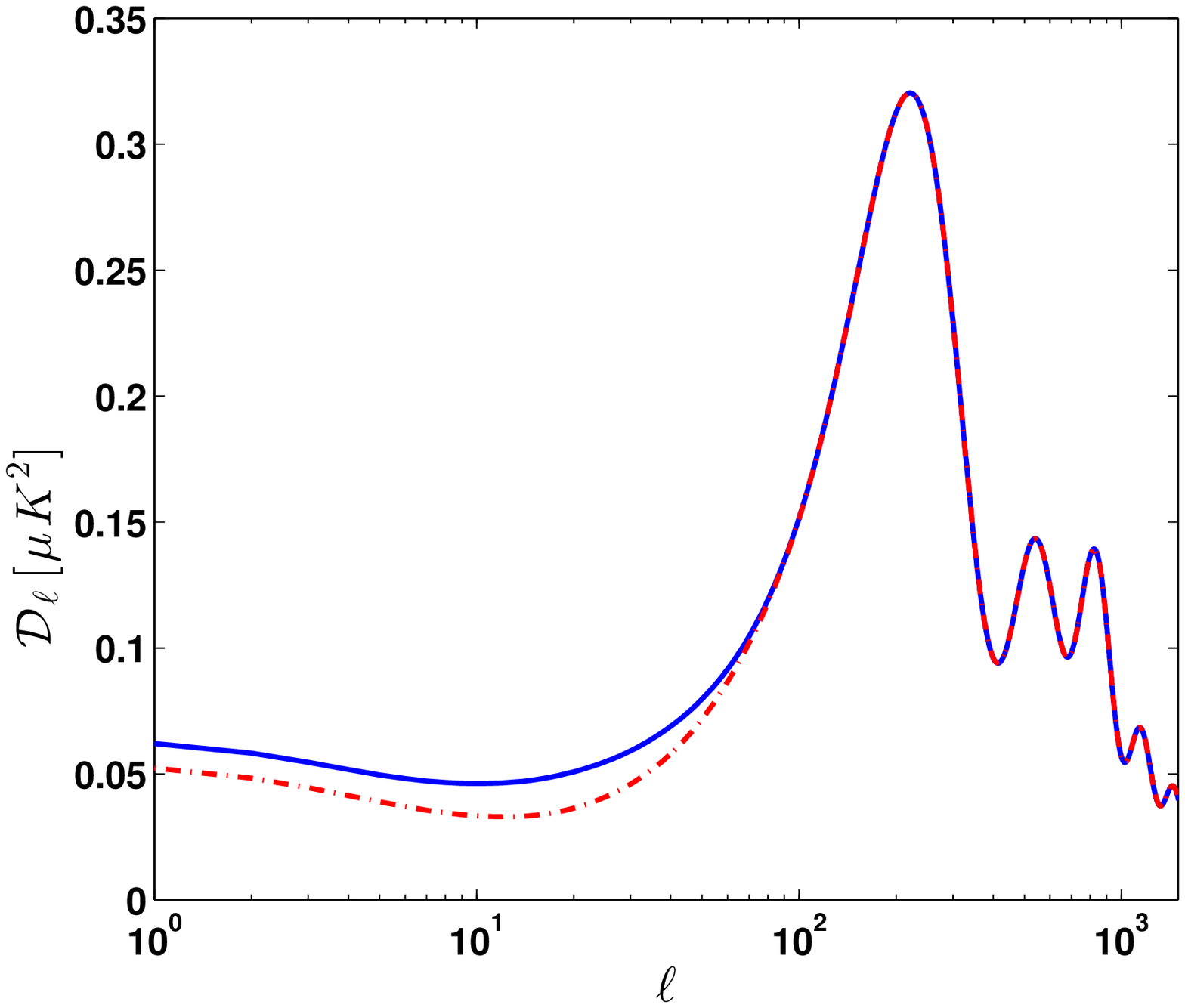}
\includegraphics[width=8cm]{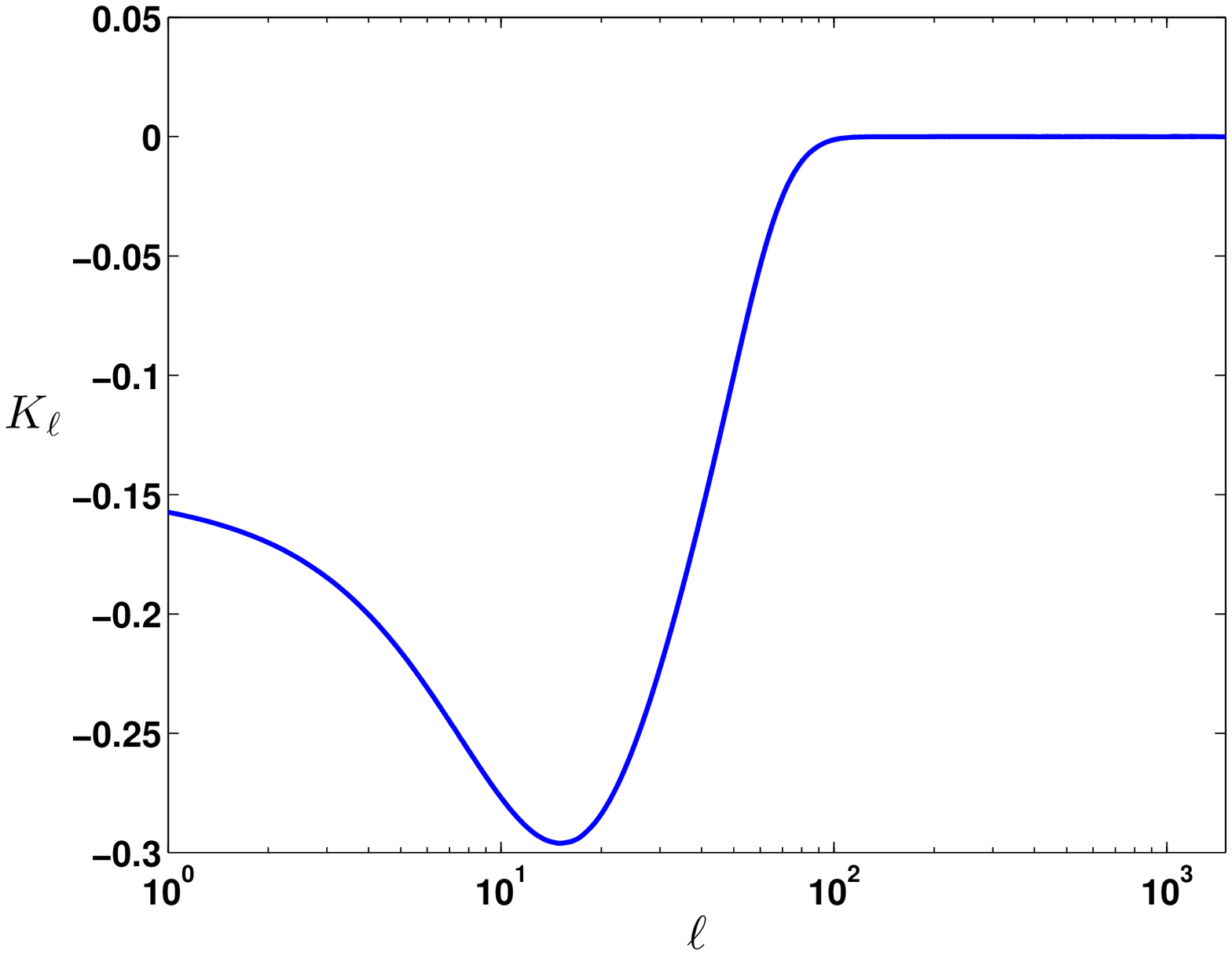}
\caption{The same plot as in Figure~\ref{case1} but for $\theta_k =\pi /4$, independent of scale. 
As expected, there is no oscillations in the amplitude of dipole asymmetry while it rapidly falls off on large $\ell$.  
Here we have  set $N_0=10$, $\bar A=-0.014$ and for the momenta in Mpc$^{-1}$ units: $k_L = 0.00006$, $k_c = 0.002$ and $k_0= 0.0000095$. The effective amplitude of asymmetry is $\calB \simeq 0.0715$.}
\label{case2}
\end{figure}

In Figure~\ref{case1} we have presented the case in which $-\eta_0=1/k_0$ is fixed so 
$\theta_k$ linearly depends on $k$. There are two important features. First, for large $k$,
the amplitude of the dipole asymmetry vanishes exponentially. This can easily account for the absence of dipole asymmetries in CMB power spectrum for large $\ell$ as discussed before. The second feature is that due to the scale-dependence of $\theta_k$ the amplitude of dipole asymmetry oscillates with scales. Both of these predictions can be tested in a careful CMB map investigation.  
In Figure~\ref{case2} we have presented the second case in which $\theta_0$ is fixed for all modes. As expected, there is no oscillation in dipole amplitude. However, as in the previous case,  the amplitude of dipole asymmetry falls off exponentially. As in previous case, these predictions for the dipole 
asymmetry can be tested  in CMB maps.

Until now we have assumed that a single super-horizon mode was invoked as the source of 
dipole asymmetry. Alternatively,  one may assume a continuum of modes in the range $k_\text{IR}< k_L< k_\text{cmb}$ all contribute
in which $k_\text{IR}$ represents the infrared cutoff of the inflationary model. In the light of~\cite{Lyth:2007jh}, $k_\text{IR}$ may be related to the size of the large box universes  $L$ via $k_\text{IR} \sim 1/L$.  Since $\calR_L$ is a random variable, $|\nabla\calR_L|$ is understood as an averaged value,
\begin{equation}
|\nabla\calR_L| = \sqrt{ \left\langle (\nabla\calR_L)^2 \right\rangle } \, ,
\end{equation}
where the average is performed over the fictitious large box where $\calR_L$ is taken. Then, this reads the first spectral moment and the rest of the analysis will go parallel to single mode modulation case. Notice, however, that to generate a sizable asymmetry we need a coherent variation of all the super-horizon modes in the large box: in general each mode will interfere each other, reducing the amplitude of the resulting asymmetry.

\section{Conclusions}
\label{sec:conc}

In this work we have considered non-vacuum initial conditions which produce large non-Gaussinity in the squeezed regime violating the non-Gaussianity consistency condition. As a result large observable hemispherical asymmetry is generated from the long mode modulation with a non-BD initial condition.

We have shown that the amplitude of the dipole asymmetry exponentially falls off on large $\ell$ which can easily address the lack of hemispherical asymmetry on $\ell >600$ as well as sub-CMB  scales. In addition, depending on the nature of the non-BD vacuum, oscillatory features show up 
in the amplitude of dipole asymmetry which can be tested in a careful CMB map study. Interestingly, the non-BD initial condition provides a natural mechanism to enhance the amplitude of the super-horizon mode compared to the CMB-scale modes. This is usually put by hand in simple models with no dynamical explanation.

It is worth mentioning  that there are additional constraints from the CMB dipole, quadrupole and octupole moments which have to be imposed in model parameters as in~\cite{Erickcek:2009at}.
These constraints usually put upper bound on the variation of the long mode within our Hubble patch, i.e. the combination $k_L x_{\mathrm{cmb}} \calP_{L}^{1/2}\,$ which appears in (\ref{AfNL}).  This means that to be able to obtain the observed hemispherical asymmetry, we need to increase $\calB$ to a large enough value. This can be easily done by reducing the sound speed since there is no tight constraint on $c_s$  in this model, as we discussed before. Having this said, an accurate calculation of low $\ell$ multipoles is beyond the scope of this article. Note that in our approach 
we have started from power spectrum, not the curvature perturbation itself, so it is not straightforward to calculate multipoles.

Final remarks are in order on the primordial tensor perturbation. 
There are evidences for the detection of the primordial gravitational waves by the BICEP2 observations~\cite{Ade:2014xna}. Interestingly, it was proposed in~\cite{Chluba:2014uba} that the hemispherical asymmetry in gravitational waves may be behind the apparent tension between the BICEP2 and PLANCK observations. This is in the light of proposal in~\cite{Abolhasani:2013vaa} in which it is 
predicted that a hemispherical asymmetry from the long mode modulation in curvature perturbation power spectrum also {\em induces} a hemispherical asymmetry in tensor perturbations. It will be very interesting to investigate the proposals from~\cite{Abolhasani:2013vaa,Chluba:2014uba}
to see whether or not the dipole asymmetry in primordial power spectra is behind the 
apparent tension between the BICEP2 and PLANCK observations (see also~\cite{Ashoorioon:2014nta} which employed the non-BD initial condition to address the tension between the BICEP2 and PLANCK observations). In addition, if the upcoming PLANCK data-release observes the 
hemispherical asymmetry in tensor mode it will provide a strong support
for the idea of long mode modulation and the concept that the observed universe itself is a part of a much larger inflationary universe which can be detected only indirectly such as from the effects of long mode modulation.

\subsection*{Acknowledgment }

We thank Ali Akbar Abolhasani, Shant Baghram and Gary Shiu for discussions. 
HF and JG thank KITPC for hospitality during the workshop 
``Cosmology after Planck" where this work was initiated. 
JG acknowledges the Max-Planck-Gesellschaft, the Korea Ministry of Education, Science and Technology, Gyeongsangbuk-Do and Pohang City for
the support of the Independent Junior Research Group at the Asia Pacific Center for Theoretical Physics.
JG is also supported by a Starting Grant through the Basic Science Research Program of the National Research Foundation of Korea (2013R1A1A1006701).

\end{document}